\documentclass[10pt,twocolumn,twoside]{IEEEtran}
\usepackage{amsmath}
\usepackage{color}
\usepackage{cases}
\usepackage{amsfonts,amsmath,amssymb}
\usepackage{mathrsfs}
\usepackage{cite}
\usepackage{graphicx}
\usepackage{url}
\usepackage{enumerate}
\usepackage{subfigure}
\usepackage{hyperref}
\usepackage{algorithm}
\usepackage{algorithmic}
\usepackage{amsmath,bm}
\usepackage{caption}
\usepackage{float}
\usepackage{stfloats}
\usepackage{multirow}
\usepackage{url}
\usepackage{breakurl}
\usepackage{txfonts}
\usepackage{booktabs}
\usepackage{threeparttable}
\usepackage{multirow}
\usepackage{fancyhdr}
\usepackage{lettrine}

\usepackage{geometry}
\DeclareMathOperator*{\argmin}{arg\,min}

\begin{document}
\title{Distance Correlation Based Brain Functional Connectivity Estimation and Non-Convex Multi-Task Learning for Developmental fMRI Studies}

\author{\mbox{Li Xiao, Biao Cai, Gang Qu, Julia M. Stephen, Tony W. Wilson},\\ \mbox{ Vince D. Calhoun,~\IEEEmembership{Fellow,~IEEE}, and Yu-Ping Wang,~\IEEEmembership{Senior Member,~IEEE}}
\thanks{This work was supported in part by NIH under Grants R01GM109068, R01MH104680, R01MH107354, R01AR059781, R01EB006841, R01EB005846, R01MH103220, R01MH116782, R01MH121101, P20GM130447, P20GM103472, and in part by NSF under Grant 1539067.}
\thanks{L. Xiao, B. Cai, G. Qu, and Y.-P. Wang are with the Department of Biomedical Engineering,
Tulane University, New Orleans, LA 70118, (e-mail: wyp@tulane.edu).}
\thanks{J. M. Stephen is with the Mind Research Network, Albuquerque, NM 87106. }
\thanks{T. W. Wilson is with the Department of Neurological Sciences, University of Nebraska Medical Center, Omaha, NE 68198.}
\thanks{V. D. Calhoun is with the Tri-Institutional Center for Translational Research in Neuroimaging and Data Science (TReNDS), Georgia State University, Georgia Institute of Technology, Emory University, Atlanta, GA 30030.}
}
\maketitle

\thispagestyle{fancy}
\fancyhead{}
\lhead{}
\lfoot{\footnotesize{This work has been submitted to the IEEE for possible publication. Copyright may be transferred without notice, after which this version may no longer be accessible.}}
\cfoot{}
\rfoot{}

\begin{abstract}
Resting-state functional magnetic resonance imaging (rs-fMRI)-derived functional connectivity patterns have been extensively utilized to delineate global functional organization of the human brain in health, development, and neuropsychiatric disorders. In this paper, we investigate how functional connectivity in males and females differs in an age prediction framework. We first estimate functional connectivity between regions-of-interest (ROIs) using distance correlation instead of Pearson's correlation. Distance correlation, as a multivariate statistical method, explores spatial relations of voxel-wise time courses within individual ROIs and measures both linear and nonlinear dependence, capturing more complex information of between-ROI interactions. Then, a novel non-convex multi-task learning (NC-MTL) model is proposed to study age-related gender differences in functional connectivity, where age prediction for each gender group is viewed as one task. Specifically, in the proposed NC-MTL model, we introduce a composite regularizer with a combination of non-convex $\ell_{2,1-2}$ and $\ell_{1-2}$ regularization terms for selecting both common and task-specific features. Finally, we validate the proposed NC-MTL model along with distance correlation based functional connectivity on rs-fMRI of the Philadelphia Neurodevelopmental Cohort for predicting ages of both genders. The experimental results demonstrate that the proposed NC-MTL model outperforms other competing MTL models in age prediction, as well as characterizing developmental gender differences in functional connectivity patterns.
\end{abstract}

\begin{IEEEkeywords}
Brain development, distance correlation, feature selection, functional connectivity, multi-task learning.
\end{IEEEkeywords}

\section{Introduction}
\lettrine[lraise=0.1, nindent=0em, slope=-.5em, lines=2]{\textbf{F}}{UNCTIONAL} magnetic resonance imaging (fMRI) is a modern neuroimaging technique
that characterizes brain function and organization through hemodynamic changes \cite{glover2011,jiansong2016,Biswal2010}. In recent decades, the fMRI-derived functional connectome has attracted a great deal of interest for providing new insights into individual variations in behavior and cognition \cite{Calhoun2,xshen2017,siyuangao2019,biaojiejie}. The connectome is defined as a network architecture of functional connectivity between brain regions-of-interest (ROIs). It facilitates the understanding of fMRI brain activation patterns, and acts like a ``fingerprint'' to distinguish individuals from the population \cite{esfinn2015,zaixucui2020,biaocai2019}.

Recently, brain developmental fMRI studies have shown that the human brain undergoes important changes of functional connectome across the lifespan \cite{fair2009,lubinwang2012,anqiqiu2015}. For instance, Fair \textit{et al.} \cite{fair2009} demonstrated that the organization of several functional modules shifts from a local anatomical emphasis in children to a more distributed architecture in young adults, which might be driven by an abundance of short-range functional connections that tend to weaken over age as well as long-range functional connections that tend to strengthen over age. Accordingly, there has been a surge in work focusing on predicting an individual's age from functional connectivity \cite{dosenbach2010,meier2012,nielsen2019}, in order to potentially aid in diagnosis and prognoses of developmental disorders and neuropsychiatric diseases. However, considering that changes of age-related functional connectivity get complicated from childhood to senescence, there still remains a challenge of understanding the developmental trajectories of brain function more accurately. In this paper, we address this challenge in two ways: 1) by refining the estimation of functional connectivity to explore the intrinsic relationships between ROIs; and 2) by developing an advanced machine learning model to handle very high-dimensional functional connectivity data.

The majority of previous developmental fMRI work is based on the conventional functional connectivity analysis, in which the Pearson's correlation between two ROI-wise time courses is computed as functional connectivity between the corresponding ROIs, and each ROI-wise time course is the average of the time courses of all constituent voxels within the ROI. Although this approach provides straightforward estimates of functional connectivity, only linear dependence between ROIs is detected, and important information on the underlying true connectivity may be lost when averaging all voxel-wise time courses within an ROI. Therefore, in this paper we utilize distance correlation \cite{szekely2007,szekely2013} to quantify functional connectivity as also studied in \cite{geerligs2016,yoo2019}, for better uncovering the complex interactions between ROIs. Different from Pearson's correlation, distance correlation is a measure of both linear and nonlinear dependence between two random vectors of arbitrary dimensions. By regarding an ROI and its constituent voxels as a random vector and the components of the vector, respectively, we can directly perform on voxel-wise time courses within each ROI to compute distance correlation between ROIs. In such a way, distance correlation based functional connectivity can preserve spatial information of all voxel-wise time courses within each ROI and improve characterization of between-ROI interactions compared with Pearson's correlation. We tested their predictive power from resting-state fMRI (rs-fMRI) of the Philadelphia Neurodevelopmental Cohort (PNC) \cite{pncdata1} for each gender group separately. The experimental results demonstrate that distance correlation based functional connectivity better predicted ages of both males and females (aged $8\!-\!22$ years old) than Pearson's correlation based functional connectivity.

Furthermore, multiple studies have documented the presence of gender differences in brain development relevant to social and behavioral domains during childhood through adolescence \cite{etchell2018,vincent2007,xnzuo2010,alarcon2015}. For example, evidences suggest that females show better verbal working memory and social cognition than males, while males perform better than females on spatial orientation and motor coordination \cite{theodore2015,xfzhu2018,rubengur2012}. Inspired by the observations in these studies, in this paper we propose a novel non-convex multi-task learning (NC-MTL) model to investigate age-related gender differences in an age prediction framework, where age prediction tasks for both genders from functional connectivity are jointly analyzed. Specifically, we consider age prediction for each gender group as one task, and select age-related common and gender-specific functional connectivity features underlying brain development. To do so, we introduce a composite of the non-convex $\ell_{2,1-2}$ and $\ell_{1-2}$ regularizers in our NC-MTL model. The two regularizers have been recently used, respectively, in \cite{yshi2018} and \cite{eesser2013,pyin2015,ylou2015}, and shown to be improved alternatives to the classical $\ell_{2,1}$ and $\ell_{1}$ regularizers widely used in previous MTL models \cite{huawang2011,tabarestani2020,braind2020,jiayuzhou2012,junwang2019,xiaoli1,xiaokehao2020}. Thus, the use of the $\ell_{2,1-2}$ term induces group sparsity for selecting common features shared by all tasks, and the use of the $\ell_{1-2}$ term enables us to select task-specific features. In addition, from a machine learning point of view, adding some proper regularization term in our NC-MTL model is beneficial to avoid over-fitting, especially in the high-dimensional feature but low sample-size scenarios. To validate the effectiveness of our NC-MTL model, we conducted multiple experiments to jointly predict ages of both genders using functional connectivity from rs-fMRI of the PNC \cite{pncdata1}. The experimental results show that our NC-MTL model significantly outperformed other previous MTL models, and can characterize the developmental gender differences in functional connectivity patterns.

The remainder of this paper is organized as follows. In Section \ref{methods}, we first introduce distance correlation and apply it to measure functional connectivity. Then, we present the proposed NC-MTL model and its optimization algorithm. In Section \ref{expem}, we provide details of the experimental results and comparisons, followed by a discussion on the discovered gender differences in functional connectivity during brain development as well as the limitations and future research directions. In Section \ref{conclusion}, we conclude this paper.

Throughout this paper, we use uppercase boldface, lowercase boldface, and normal italic letters to denote matrices, vectors, and scalars, respectively. The superscript $T$ denotes the matrix transpose. $\langle\textbf{A},\textbf{B}\rangle$ stands for the inner product of two matrices $\textbf{A}$ and $\textbf{B}$, and equals the trace of $\textbf{A}^T\textbf{B}$. Let $\mathbb{R}$ denote the set of real numbers. For the sake of clarity, we summarize the frequently used notations and corresponding descriptions in Table \ref{table1}.

\begin{table}[!h]
\centering
\renewcommand\arraystretch{1}
\caption{Notations and descriptions.}
\label{table1}
\begin{tabular}{ll}
\hline
\specialrule{0em}{1.25pt}{1.25pt}
Notation & Description\\ \hline \specialrule{0em}{1.25pt}{1.25pt}
$W_{ij}$     & The $(i,j)$-th element of a matrix $\textbf{W}$.\vspace{0.05cm}\\
$\textbf{w}_i$     & The $i$-th column of a matrix $\textbf{W}$.\vspace{0.05cm}\\
$\textbf{w}^i$  & The $i$-th row of a matrix $\textbf{W}$.\vspace{0.05cm}\\
$w_i$   & The $i$-th element of a vector $\textbf{w}$.\vspace{0.05cm}\\
$\partial f$  & The set of sub-gradients of a function $f$.\vspace{0.05cm}\\
$\nabla f$   & The gradient of a differentiable function $f$.\vspace{0.05cm}\\
$\ell_p$    & $\lVert\textbf{w}\rVert_p=(\sum_i|w_i|^p)^{1/p}$ or $\lVert\textbf{W}\rVert_p=(\sum_{i,j}|W_{ij}|^p)^{1/p}$.\vspace{0.05cm}\\
$\ell_{2,p}$     & $\lVert\textbf{W}\rVert_{2,p}=(\sum_i\lVert\textbf{w}^i\lVert_2^p)^{1/p}$, and $\lVert\textbf{W}\rVert_{2,2}=\lVert\textbf{W}\rVert_2$.\vspace{0.05cm}\\
$\lVert\textbf{W}\rVert_{F}$     & The Frobenius norm of a matrix $\textbf{W}$, and $\lVert\textbf{W}\rVert_{F}=\lVert\textbf{W}\rVert_{2,2}$.\vspace{0.05cm}\\
$\textbf{W}^{(k)},\textbf{w}^{(k)},w^{(k)}$     & $\textbf{W},\textbf{w},w$ at the $k$-th iteration in an iterative algorithm.\vspace{0.05cm}\\
\specialrule{0em}{0.15pt}{0.15pt}\hline
\end{tabular}
\end{table}

\section{Methods}\label{methods}
In this section, we first briefly introduce distance correlation \cite{szekely2007,szekely2013}, and compare it with Pearson's correlation in terms of application for measuring functional connectivity. Afterwards, we propose an innovative non-convex multi-task learning (NC-MTL) model as well as its optimization algorithm. At the end, we validate the proposed NC-MTL model on synthetic data.

\subsection{Functional connectivity measured by distance correlation}\label{subsec2.1}
In contrast with Pearson's correlation, which is a widely used measure of linear dependence between two random variables, distance correlation has recently been proposed for measuring and testing general (i.e., both linear and nonlinear) dependence between two random vectors of arbitrary dimensions. Two random vectors are independent if and only if the distance correlation between them is zero \cite{szekely2007}. However, we cannot say that two random variables with Pearson's correlation being zero are independent, because they are very likely to be nonlinearly dependent. Hence, distance correlation can generally capture more complex relationships than Pearson's correlation.

Let $\{\textbf{a}_i\}_{i=1}^n$ and $\{\textbf{b}_i\}_{i=1}^n$ be $n$ paired samples from two random vectors $\textbf{a}\in\mathbb{R}^{p}$ and $\textbf{b}\in\mathbb{R}^{q}$, where the dimensions $p$ and $q$ are arbitrarily large and not necessarily required to be equal. The unbiased (sample) distance correlation between $\textbf{a}$ and $\textbf{b}$ is then defined as follows \cite{szekely2013}.

\begin{itemize}
  \item[$1)$] Calculate the Euclidean distance matrices $\textbf{A}\in\mathbb{R}^{n\times n}$ and $\textbf{B}\in\mathbb{R}^{n\times n}$ whose elements are $A_{ij}=\lVert\textbf{a}_i-\textbf{a}_j\rVert_2$ and $B_{ij}=\lVert\textbf{b}_i-\textbf{b}_j\rVert_2$ for $1\leq i,j\leq n$, respectively.
  \item[$2)$] Calculate the U-centered distance matrices $\widehat{\textbf{A}}\in\mathbb{R}^{n\times n}$ with
  \begin{equation}
  \widehat{A}_{ij}=
  \begin{cases}
  A_{ij}-\frac{\sum_{l=1}^{n}A_{il}}{n-2}-\frac{\sum_{k=1}^{n}A_{kj}}{n-2}+\frac{\sum_{k,l=1}^{n}A_{kl}}{(n-1)(n-2)}, & i\neq j,\\
  0, & i=j,
  \end{cases}
  \end{equation}
  for $1\leq i,j\leq n$ and $\widehat{\textbf{B}}\in\mathbb{R}^{n\times n}$ accordingly.
  \item[$3)$] Define the distance covariance (dCov) by
  \begin{equation}
  \text{dCov}(\textbf{a},\textbf{b})=\frac{\sum_{i\neq j}\widehat{A}_{ij}\widehat{B}_{ij}}{n(n-3)}.
  \end{equation}
  \item[$4)$] Define the distance correlation (dCor) by
  \begin{equation}
  \text{dCor}(\textbf{a},\textbf{b})=
  \sqrt{\frac{\text{dCov}(\textbf{a},\textbf{b})}{\sqrt{\text{dCov}(\textbf{a},\textbf{a})\text{dCov}(\textbf{b},\textbf{b})}}}
  \end{equation}
  if $\text{dCov}(\textbf{a},\textbf{b})>0$, and otherwise $0$.
\end{itemize}

\begin{figure}[!t]
  \centering
\includegraphics[width=0.9\columnwidth]{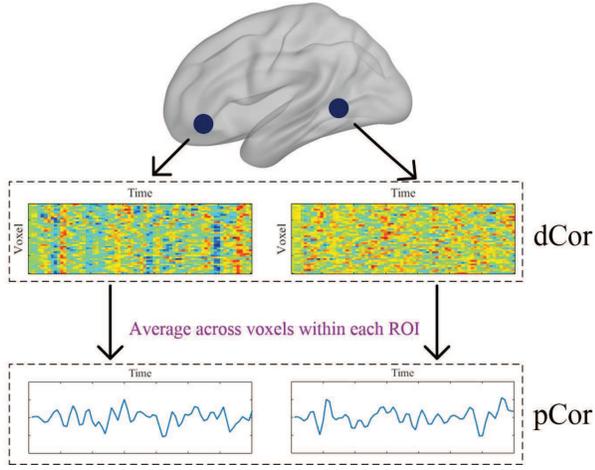}\\
  \caption{An illustration of the difference between dCor based functional connectivity and pCor based functional connectivity. At the top, each blue dot denotes an ROI; in the middle, each heatmap shows all voxel-wise time courses within the corresponding ROI; at the bottom, each line plot represents an ROI-wise time course calculated by averaging all voxel-wise time courses within the corresponding ROI.}\label{fig1}
\end{figure}

Without loss of generality, by regarding $\textbf{a}$ and $\textbf{b}$ as a pair of ROIs consisting of $p$ and $q$ voxels, respectively, and $\{\textbf{a}_i\}_{i=1}^n$ and $\{\textbf{b}_i\}_{i=1}^n$ as the corresponding voxel-wise time courses within them over a total of $n$ time points, we can compute the distance correlation, i.e., $\text{dCor}(\textbf{a},\textbf{b})$, to quantify functional connectivity between them \cite{geerligs2016,yoo2019}. As all voxel-wise time courses within an ROI are utilized by treating each voxel as one variable, dCor is a multivariate measure of functional connectivity. By comparison, Pearson's correlation (pCor) is a univariate measure of functional connectivity, where each ROI is first reduced to one dimension by averaging voxel-wise time courses within it to yield one ROI-wise time course, and then functional connectivity between a pair of ROIs is measured by the pCor between their ROI-wise time courses. The difference between the two functional connectivity methods is illustrated in Fig. \ref{fig1}. It has been demonstrated in \cite{geerligs2016,yoo2019} that dCor based functional connectivity is capable of preserving the voxel-level information, resulting in improved characterization of between-ROI interactions, while averaging all voxel-wise time courses within each ROI in pCor based functional connectivity might lose important information on the underlying true connectivity. Of note, ``univariate'' and ``multivariate'' here are used to refer to the number of variables within an ROI \cite{geerligs2016}.

\subsection{Novel non-convex multi-task learning (NC-MTL)}
We assume that there are $M$ learning tasks for the data in a $d$-dimensional feature space. In the $i$-th task for $1\leq i\leq M$, we have a training dataset $\{\textbf{X}_i,\textbf{y}_i\}$, where $\textbf{X}_i\in\mathbb{R}^{n_i\times d}$ is the data matrix with $n_i$ training subjects as row vectors, each consisting of $d$ features, and $\textbf{y}_i\in\mathbb{R}^{n_i}$ is the corresponding label vector. Let $\textbf{w}_i\in\mathbb{R}^{d}$ denote the weights of all features to linearly regress the labels $\textbf{y}_i$ on $\textbf{X}_i$ in the $i$-th task. Then, an MTL model for the data can be formulated by the following optimization problem:
\begin{equation}\label{general}
\min_{\textbf{W}} \sum_{i=1}^{M}\frac{1}{2}\lVert\textbf{y}_i-\textbf{X}_i\textbf{w}_i\rVert_2^2+\alpha\Omega(\textbf{W}),
\end{equation}
where $\textbf{W}=[\textbf{w}_1,\textbf{w}_2,\cdots,\textbf{w}_M]\in\mathbb{R}^{d\times M}$ is the weight matrix of features on all tasks, $\Omega(\textbf{W})$ is the sparsity regularizer imposed for feature selection, and $\alpha>0$ is the regularization parameter that balances the tradeoff between residual error and sparsity. Through solving (\ref{general}), we obtain a sparse weight matrix $\textbf{W}^{\ast}$ to evaluate the relationship between features and labels, thereby selecting the most discriminative features across all tasks. Note that if the number of tasks equals $1$, i.e., $M=1$, then $\textbf{W}=\textbf{w}_1\in\mathbb{R}^d$ becomes the weight vector on one task, and (\ref{general}) represents single-task learning (STL).

A classical MTL model is to select common features shared by all tasks based on a group sparsity regularizer, i.e.,  $\Omega(\textbf{W})=\lVert\textbf{W}\rVert_{2,0}$, in (\ref{general}). The $\ell_{2,0}$ regularizer, extending the $\ell_{0}$ regularizer in STL to MTL, penalizes every row of $\textbf{W}$ as a whole, and enforces sparsity among the rows. As the $\ell_{2,0}$ regularizer leads to a combinatorially NP-hard optimization problem, its several approximations, such as the $\ell_{2,p}$ regularizer ($\lVert\textbf{W}\rVert_{2,p}$) with $0<p\leq1$, have been studied. Remarkably, the $\ell_{2,1}$ regularizer has been proposed as a convex approximation to the $\ell_{2,0}$ regularizer \cite{argyriou2007,nie2010,zhu2016}, and MTL in (\ref{general}) becomes
\begin{equation}
\min_{\textbf{W}} \sum_{i=1}^{M}\frac{1}{2}\lVert\textbf{y}_i-\textbf{X}_i\textbf{w}_i\rVert_2^2+\alpha\lVert\textbf{W}\rVert_{2,1},
\end{equation}
which performs well and can be easily optimized. On the other hand, as $\ell_{2,p}$ with $0<p<1$ is geometrically much closer to $\ell_{2,0}$ than $\ell_{2,1}$, the $\ell_{2,p}$ regularizer with $0<p<1$ has been developed and theoretically proven to outperform the $\ell_{2,1}$ regularizer for feature selection \cite{mzhang2014,hpeng2017,xdu2016}. However, due to the non-convexity and non-Lipschitz continuity of the $\ell_{2,p}$ regularizer with $0<p<1$, it is more challenging to solve the optimization problem in MTL. To this end, the non-convex but Lipschitz continuous $\ell_{2,1-2}$ regularizer has recently been investigated in \cite{yshi2018}, which extends the $\ell_{1-2}$ regularizer in STL \cite{eesser2013,pyin2015,ylou2015} to MTL, i.e.,
\begin{equation}
\min_{\textbf{W}} \sum_{i=1}^{M}\frac{1}{2}\lVert\textbf{y}_i-\textbf{X}_i\textbf{w}_i\rVert_2^2+\alpha\lVert\textbf{W}\rVert_{2,1-2},
\end{equation}
where $\lVert\textbf{W}\rVert_{2,1-2}\triangleq\lVert\textbf{W}\rVert_{2,1}-\lVert\textbf{W}\rVert_{2,2}=\lVert\textbf{W}\rVert_{2,1}-\lVert\textbf{W}\rVert_{F}$ and it is ready to verify $\lVert\textbf{W}\rVert_{2,1-2}\geq0$ due to $\lVert\textbf{W}\rVert_{F}\leq\lVert\textbf{W}\rVert_{2,1}$.
The $\ell_{2,1-2}$ regularizer has been shown to not only achieve better feature selection performance, but also result in an easier optimization problem because of the non-Lipschitz continuity.

\begin{figure}[!t]
  \centering
\includegraphics[width=0.9\columnwidth]{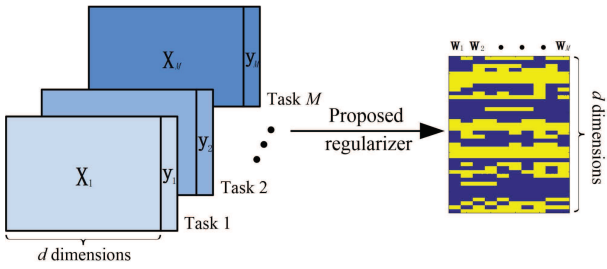}\\
  \caption{An illustration of the proposed NC-MTL model in (\ref{qingtian2}). The left-hand side shows the input datasets $\{\textbf{X}_i,\textbf{y}_i\}_{i=1}^M$, and the right-hand side shows the sparsity pattern of the learned weight matrix $\textbf{W}$.}\label{fig2}
\end{figure}

As we mentioned above, all of the $\ell_{2,p}$ with $0<p\leq1$ and $\ell_{2,1-2}$ regularizers are approximations to the $\ell_{2,0}$ regularizer in MTL. So, they can achieve the group sparsity and only select common features shared by all tasks, but fail to consider task-specific features (i.e., features shared by a subset of tasks). To extract both common and task-specific features in MTL, we introduce a composite of the $\ell_{2,1-2}$ and $\ell_{1-2}$ regularizers, and obtain the following NC-MTL model
\begin{equation}\label{qingtian}
\min_{\textbf{W}} \sum_{i=1}^{M}\frac{1}{2}\lVert\textbf{y}_i-\textbf{X}_i\textbf{w}_i\rVert_2^2+\alpha\lVert\textbf{W}\rVert_{2,1-2}+\beta\lVert\textbf{W}\rVert_{1-2},
\end{equation}
i.e.,
\begin{equation}\label{qingtian2}
\min_{\textbf{W}} \sum_{i=1}^{M}\frac{1}{2}\lVert\textbf{y}_i-\textbf{X}_i\textbf{w}_i\rVert_2^2+\alpha\lVert\textbf{W}\rVert_{2,1}+\beta\lVert\textbf{W}\rVert_{1}-(\alpha+\beta)\lVert\textbf{W}\rVert_{F},
\end{equation}
where $\lVert\textbf{W}\rVert_{1-2}\triangleq\lVert\textbf{W}\rVert_{1}-\lVert\textbf{W}\rVert_{F}$ is used to enforce the sparsity among all elements in $\textbf{W}$ and we immediately have $\lVert\textbf{W}\rVert_{1-2}\geq0$ due to $\lVert\textbf{W}\rVert_{F}\leq\lVert\textbf{W}\rVert_{1}$. It is worth noting that, the first term $\ell_{2,1-2}$ of the composite regularizer in (\ref{qingtian}) achieves the group sparsity to select common features shared by all tasks, while the second term $\ell_{1-2}$ contributes to selecting task-specific features. The two terms are improved alternatives to $\ell_{2,1}$ and $\ell_{1}$ respectively, which have been used in several existing MTL models (see, e.g., \cite{huawang2011,tabarestani2020,braind2020,jiayuzhou2012,junwang2019,xiaoli1,xiaokehao2020}). Hyperparameters $\alpha, \beta>0$ control the balance between the sparsity patterns of common and task-specific features. The illustration of the proposed NC-MTL model is shown in Fig. \ref{fig2}.

\subsection{Optimization algorithm for NC-MTL}
Let us consider the proposed NC-MTL model in (\ref{qingtian2}), whose objective function, denoted as $h(\textbf{W})$, is non-convex and the subtraction of two convex functions $f(\textbf{W})$ and $g(\textbf{W})$, i.e.,
\begin{equation}\label{chouxiang}
\min_{\textbf{W}}\; h(\textbf{W}):= f(\textbf{W})-g(\textbf{W})
\end{equation}
with
\begin{equation}
f(\textbf{W})=\sum_{i=1}^{M}\frac{1}{2}\lVert\textbf{y}_i-\textbf{X}_i\textbf{w}_i\rVert_2^2+\alpha\lVert\textbf{W}\rVert_{2,1}+\beta\lVert\textbf{W}\rVert_{1} \;\text{ and }
\end{equation}
\begin{equation}
g(\textbf{W})=(\alpha+\beta)\lVert\textbf{W}\rVert_{F}.
\end{equation}
A well-known scheme for addressing such a non-convex optimization problem is first to linearize $g(\textbf{W})$ using its 1st-order Taylor-series expansion at the current solution $\textbf{W}^{(k)}$, and then advance to a new one $\textbf{W}^{(k+1)}$ by solving a convex optimization subproblem in the framework of ConCave-Convex Procedure (CCCP) \cite{yuille2003}.

More specifically, the CCCP algorithm can solve the above problem (\ref{chouxiang}) with the following iterations.
\begin{equation}\label{iteal}
\begin{split}
\textbf{W}^{(k+1)}&=\argmin_{\textbf{W}} f(\textbf{W})-\left(g(\textbf{W}^{(k)})+\langle\textbf{W}-\textbf{W}^{(k)},\textbf{S}^{(k)}\rangle\right)\\
&=\argmin_{\textbf{W}} f(\textbf{W})-\langle\textbf{W},\textbf{S}^{(k)}\rangle,
\end{split}
\end{equation}
where $\textbf{S}^{(k)}\in\partial g(\textbf{W}^{(k)})$. Following the definition of sub-gradient, i.e., for any $\textbf{W}$,
$g(\textbf{W})\geq g(\textbf{W}^{(k)})+\langle\textbf{W}-\textbf{W}^{(k)},\textbf{S}^{(k)}\rangle$, we obtain
\begin{equation}
\begin{split}
h(\textbf{W}^{(k)})&=f(\textbf{W}^{(k)})-g(\textbf{W}^{(k)})\\
&\geq f(\textbf{W}^{(k+1)})-\left(g(\textbf{W}^{(k)})+\langle\textbf{W}^{(k+1)}-\textbf{W}^{(k)},\textbf{S}^{(k)}\rangle\right)\\
&\geq f(\textbf{W}^{(k+1)})-g(\textbf{W}^{(k+1)})=h(\textbf{W}^{(k+1)}).
\end{split}
\end{equation}
Therefore, the objective function values $\{h(\textbf{W}^{(k)})\}_{k=0}^{\infty}$ are monotonically decreasing. Moreover, from the formula of the objective function $h(\textbf{W})$ in (\ref{qingtian2}), $\{h(\textbf{W}^{(k)})\}_{k=0}^{\infty}$ are bounded below by zero, and they thus converge. We can obtain a local optimal $\textbf{W}^{\star}$ of (\ref{qingtian2}) by iteratively solving (\ref{iteal}); see Algorithm \ref{alg1} for details.

\begin{algorithm}[!t]
\caption{CCCP for solving the proposed NC-MTL in (\ref{qingtian2})}

\vspace*{0.4mm}

\textbf{Input:} Datasets $\{\textbf{X}_i,\textbf{y}_i\}_{i=1}^{M}$; hyperparameters $\alpha,\beta>0$.\\
\vspace{-0.4cm}
\begin{algorithmic}[1]\label{alg1}
\STATE \textbf{Initialize} $k=0$ and $\textbf{W}^{(0)}=\textbf{0}$;
\STATE \textbf{repeat}
\STATE \quad $\textbf{W}^{(k+1)}:=$
\begin{equation}\label{subpro2}
\argmin_{\textbf{W}} \sum_{i=1}^{M}\frac{1}{2}\lVert\textbf{y}_i-\textbf{X}_i\textbf{w}_i\rVert_2^2+\alpha\lVert\textbf{W}\rVert_{2,1}+\beta\lVert\textbf{W}\rVert_{1}-\langle\textbf{W},\textbf{S}^{(k)}\rangle,
\end{equation}
\quad where $\textbf{S}^{(k)}\in\partial g(\textbf{W}^{(k)})$ is taken as
\begin{equation}
\textbf{S}^{(k)}=
\begin{cases}
  (\alpha+\beta)\lVert\textbf{W}^{(k)}\rVert_{F}^{-1}\textbf{W}^{(k)}, & \textbf{W}^{(k)}\neq\textbf{0},\\
  \textbf{0}, & \textbf{W}^{(k)}=\textbf{0};
  \end{cases}
\end{equation}
\STATE \quad $k:=k+1$;
\STATE \textbf{until} convergence.
\end{algorithmic}
\textbf{Output:} The optimal solution $\textbf{W}^{\star}$.
\end{algorithm}

We next use the accelerated proximal gradient (APG) algorithm \cite{nesterov1983} to solve the convex subproblem (\ref{iteal}) or (\ref{subpro2}), whose objective function is the summation of two convex functions, i.e., $\phi(\textbf{W})$ (differentiable) and $\varphi(\textbf{W})$ (non-differentiable) with
\begin{equation}
\phi(\textbf{W})=\sum_{i=1}^{M}\frac{1}{2}\lVert\textbf{y}_i-\textbf{X}_i\textbf{w}_i\rVert_2^2-\langle\textbf{W},\textbf{S}^{(k)}\rangle \;\text{ and }
\end{equation}
\begin{equation}
\varphi(\textbf{W})=\alpha\lVert\textbf{W}\rVert_{2,1}+\beta\lVert\textbf{W}\rVert_{1}.
\end{equation}
Specifically, we iteratively update $\textbf{W}$ as follows.
\begin{equation}\label{aada}
\textbf{W}^{(t+1)}=\argmin_{\textbf{W}}\Lambda_l(\textbf{W},\textbf{W}^{(t)}),
\end{equation}
where $\Lambda_l(\textbf{W},\textbf{W}^{(t)})=\phi(\textbf{W}^{(t)})+\langle\textbf{W}-\textbf{W}^{(t)},\nabla\phi(\textbf{W}^{(t)})\rangle+\frac{1}{2l}\lVert\textbf{W}-\textbf{W}^{(t)}\rVert_{F}^2+\varphi(\textbf{W})$, and $l$ is a variable step size. In matrix calculus, the gradient of a scalar-valued function $\phi(\textbf{W})$ with respect to $\textbf{W}$ can be written as a vector whose components are the gradients of $\phi$ with respect to every column of $\textbf{W}$. Therefore, we obtain
$\nabla\phi(\textbf{W}^{(t)})=[\nabla\phi(\textbf{w}_1^{(t)}),\nabla\phi(\textbf{w}_2^{(t)}),\cdots,\nabla\phi(\textbf{w}_M^{(t)})]$, and $\nabla\phi(\textbf{w}_i^{(t)})$ for $1\leq i\leq M$ can be easily calculated as
\begin{equation}
\nabla\phi(\textbf{w}_i^{(t)})=\textbf{X}_i^{T}(\textbf{X}_i\textbf{w}_i^{(t)}-\textbf{y}_i)-\textbf{s}_i^{(k)},
\end{equation}
where $\textbf{w}_i^{(t)}$ and $\textbf{s}_i^{(k)}$ represent the $i$-th columns of $\textbf{W}^{(t)}$ and $\textbf{S}^{(k)}$, respectively.
Based on simple calculation, we can equivalently rewrite $\Lambda_l(\textbf{W},\textbf{W}^{(t)})$ as
$\Lambda_l(\textbf{W},\textbf{W}^{(t)})=\phi(\textbf{W}^{(t)})-\frac{l}{2}\lVert\nabla\phi(\textbf{W}^{(t)})\rVert_{F}^{2}+\frac{1}{2l}\lVert\textbf{W}-\textbf{W}^{(t)}+l\nabla\phi(\textbf{W}^{(t)})\rVert_{F}^{2}+\varphi(\textbf{W})$.
Then, after ignoring the items independent of $\textbf{W}$ in (\ref{aada}), the update procedure becomes
\begin{equation}\label{optim}
\textbf{W}^{(t+1)}=\argmin_{\textbf{W}}\frac{1}{2}\lVert\textbf{W}-\textbf{V}^{(t)}\rVert_{F}^{2}+l\varphi(\textbf{W}),
\end{equation}
where $\textbf{V}^{(t)}=\textbf{W}^{(t)}-l\nabla\phi(\textbf{W}^{(t)})$. Clearly, (\ref{optim}) is in fact,
\begin{equation}
\textbf{W}^{(t+1)}=\text{prox}_{l\varphi}(\textbf{V}^{(t)}),
\end{equation}
where $\text{prox}_{l\varphi}$ stands for the proximal operator \cite{nparikh2014} of the scaled function $l\varphi$.

Owing to the separability of $\textbf{W}$ on its rows in (\ref{optim}), we can decouple (\ref{optim}) into the following optimization problem for each row independently, i.e., for $1\leq i\leq d$,
\begin{equation}\label{eqdong}
\begin{split}
\textbf{w}^{(t+1),i}&=\argmin_{\textbf{w}^i}\frac{1}{2}\lVert\textbf{w}^i-\textbf{v}^{(t),i}\rVert_{2}^{2}+l\alpha\lVert\textbf{w}^i\rVert_2+l\beta\lVert\textbf{w}^i\rVert_1\\
&=\text{prox}_{l\tau}(\textbf{v}^{(t),i}),
\end{split}
\end{equation}
where $\textbf{w}^{(t+1),i},\textbf{w}^i$, and $\textbf{v}^{(t),i}$ represent the $i$-th rows of $\textbf{W}^{(t+1)},\textbf{W}$, and $\textbf{V}^{(t)}$, respectively, and $\tau(\textbf{w}^i)=\alpha\lVert\textbf{w}^i\rVert_2+\beta\lVert\textbf{w}^i\rVert_1$ is a function of vector $\textbf{w}^i$. Letting $\tau_1(\textbf{w}^i)=\beta\lVert\textbf{w}^i\rVert_1$ and $\tau_2(\textbf{w}^i)=\alpha\lVert\textbf{w}^i\rVert_2$, we have, from \cite{jiayuzhou2012}, $\text{prox}_{l\tau}(\textbf{v}^{(t),i})=\text{prox}_{l\tau_2}(\text{prox}_{l\tau_1}(\textbf{v}^{(t),i}))$.
It is well known that both $\text{prox}_{l\tau_1}$ and $\text{prox}_{l\tau_2}$ have closed-form solutions \cite{nparikh2014}, i.e., $\textbf{r}=\text{prox}_{l\tau_1}(\textbf{u})$ with
\begin{equation}
r_i=
\begin{cases}
\left(1-\frac{l\beta}{|u_i|}\right)u_i, & \text{if }|u_i|\geq l\beta,\\
0, & \text{otherwise},
\end{cases}
\end{equation}
where $r_i$ and $u_i$ represent the $i$-th elements of vectors $\textbf{r}$ and $\textbf{u}$, respectively, and
\begin{equation}\label{eqxi}
\text{prox}_{l\tau_2}(\textbf{u})=
\begin{cases}
  \left(1-\frac{l\alpha}{\lVert\textbf{u}\rVert_2}\right)\textbf{u}, & \text{if }\lVert\textbf{u}\rVert_2\geq l\alpha,\\
  \textbf{0}, & \text{otherwise}.
\end{cases}
\end{equation}
Therefore, based on (\ref{eqdong})--(\ref{eqxi}), we can obtain the closed-form solution of $\textbf{W}^{(t+1)}$ in (\ref{optim}). To accelerate the proximal gradient method, we introduce an auxiliary variable as
\begin{equation}\label{jisuanQ}
\textbf{Q}^{(t)}=\textbf{W}^{(t)}+\frac{\theta^{(t-1)}-1}{\theta^{(t)}}(\textbf{W}^{(t)}-\textbf{W}^{(t-1)}),
\end{equation}
and perform the gradient descent procedure with respect to $\textbf{Q}^{(t)}$ instead of $\textbf{W}^{(t)}$, where the coefficient $\theta^{(t)}$ is updated by
\begin{equation}
\theta^{(t)}=\frac{1+\sqrt{1+4(\theta^{(t-1)})^2}}{2}.
\end{equation}

The pseudo-code of the APG algorithm for solving (\ref{subpro2}) is shown in Algorithm \ref{alg2}.

\begin{algorithm}[!t]
\caption{APG for solving the subproblem in (\ref{subpro2})}

\vspace*{0.4mm}

\textbf{Input:} Datasets $\{\textbf{X}_i,\textbf{y}_i\}_{i=1}^{M}$; hyperparameters $\alpha,\beta>0$.\\
\vspace{-0.4cm}
\begin{algorithmic}[1]\label{alg2}
\STATE \textbf{Initialize} $t=1,\theta^{(0)}=1,l_0=1,\sigma=0.5,\textbf{W}^{(0)}=\textbf{W}^{(1)}=\textbf{0}$;
\STATE \textbf{repeat}
\STATE \quad calculate $\textbf{Q}^{(t)}$ by (\ref{jisuanQ});
\STATE \quad $l=l_{t-1}$;
\STATE \quad\quad \textbf{while }$\phi(\textbf{W}^{(t+1)})+\varphi(\textbf{W}^{(t+1)})>\Lambda_l(\textbf{W}^{(t+1)},\textbf{Q}^{(t)})$,
      where \hspace*{0.75cm}$\textbf{W}^{(t+1)}$ is calculated by (\ref{optim}), \textbf{do}
\STATE \quad\quad\quad $l=\sigma l$;
\STATE \quad\quad \textbf{end while}
\STATE \quad $l_t=l$;
\STATE \quad $t:=t+1$;
\STATE \textbf{until} convergence.
\end{algorithmic}
\textbf{Output:} The optimal solution $\textbf{W}^{\star}$.
\end{algorithm}

\subsection{Testing the proposed NC-MTL on synthetic data}
We demonstrate the effectiveness of the proposed NC-MTL model in (\ref{qingtian2}) first on synthetic data through a comparison with other competing MTL models. We simulated a dataset with $M=10$ tasks and $d=100$ features, and each task has $40$ samples. We randomly selected $6$ features as common features shared by all $10$ tasks and $4$ features as task-specific features for each task. The weights of the selected features were generated from the uniform distribution $\mathcal{U}(1,3)$ and the weights of the remaining features were zero (see Fig. \ref{weight}(a)). The elements of the inputs $\textbf{X}_i\in\mathbb{R}^{40\times 100}$ for $1\leq i\leq10$ were generated from the Gaussian distribution $\mathcal{N}(0,2)$, and the corresponding label vectors $\textbf{y}_i\in\mathbb{R}^{40}$ were calculated as $\textbf{y}_i=\textbf{X}_i\textbf{w}_i+\bm{\epsilon}_i$, in which the elements of noise vectors $\bm{\epsilon}_i\in\mathbb{R}^{40}$ were generated from $\mathcal{N}(0,0.1)$.

\begin{figure}[!t]
 \centering
\includegraphics[width=0.9\columnwidth]{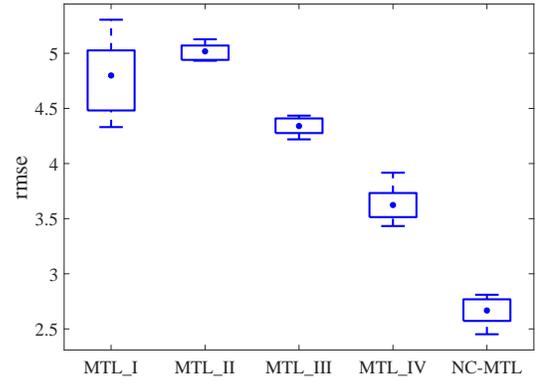}\\
\vspace{-0.25cm}
\caption{Comparison of the rmse performance of all five MTL models, where box plots show the rmse results with the error bars representing the 25-th and 75-th percentiles, respectively, and the mean values are indicated by $\bullet$. }\label{sim_result}
\end{figure}

Based on the simulated data, we compared the performance of our NC-MTL model and the following four popular MTL models.
\begin{itemize}
  \item [$1)$] MTL\_\uppercase\expandafter{\romannumeral1}: The model utilizes the $\ell_1$ regularizer to enforce feature sparsity in MTL, i.e., $\Omega(\textbf{W})=\lVert\textbf{W}\rVert_{1}$ in (\ref{general}), which is Lasso in MTL with all tasks sharing the same sparsity parameter.
  \item [$2)$] MTL\_\uppercase\expandafter{\romannumeral2} \cite{argyriou2007}: In the model, the $\ell_{2,1}$ regularizer is used to induce the group sparsity in MTL, i.e., $\Omega(\textbf{W})=\lVert\textbf{W}\rVert_{2,1}$ in (\ref{general}), for selecting common features shared by all tasks.
  \item [$3)$] MTL\_\uppercase\expandafter{\romannumeral3} \cite{yshi2018}: The model applies the $\ell_{2,1-2}$ regularizer in MTL, i.e., $\Omega(\textbf{W})=\lVert\textbf{W}\rVert_{2,1-2}$ in (\ref{general}), which is an improved alternative to the $\ell_{2,1}$ regularizer for feature selection.
  \item [$4)$] MTL\_\uppercase\expandafter{\romannumeral4} \cite{huawang2011}: In the model, the $\ell_{2,1}$ and $\ell_1$ regularizers are adopted in MTL, i.e., $\Omega(\textbf{W})=\lVert\textbf{W}\rVert_{2,1}+\frac{\beta}{\alpha}\lVert\textbf{W}\rVert_{1}$ in (\ref{general}), to select common and task-specific features, respectively.
\end{itemize}

In Fig. \ref{sim_result}, we present the average prediction performance of the five MTL models, which was quantified using root mean square error (rmse) for all the test samples of $10$ tasks over $10$ times $5$-fold nested cross-validation (CV). The regularization parameters in the MTL models were tuned from the range of $\{0.1,0.5,1,5,10,$ $50,100,150,200,250,300\}$. In Fig. \ref{weight}(b)-(f), the average of the learned weight matrices over all runs of CV is shown for each MTL model. We can observe from Figs. \ref{sim_result} and \ref{weight} that the proposed NC-MTL model extracted the most accurate features and achieved the best performance.

\begin{figure}[!t]
 \centering
\hspace*{-0.7cm}
\includegraphics[width=1.13\columnwidth]{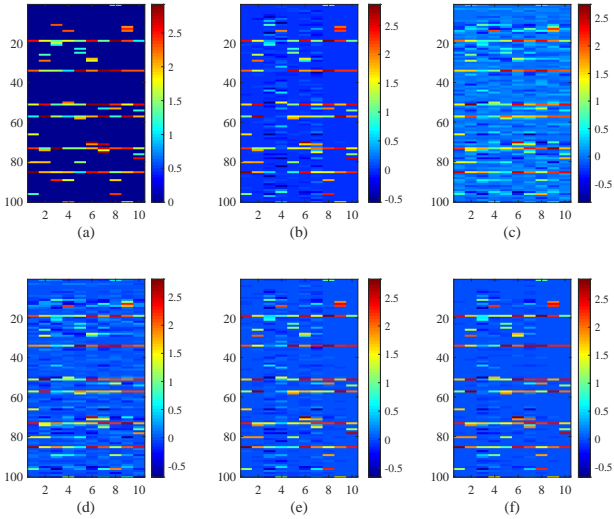}\\
\vspace{-0.25cm}
\caption{(a) The ground-truth weight matrix $\textbf{W}\in\mathbb{R}^{100\times10}$. (b)-(f) The average of the learned weight matrices over all runs of CV for each of the five MTL models (i.e., MTL\_\uppercase\expandafter{\romannumeral1}, MTL\_\uppercase\expandafter{\romannumeral2}, MTL\_\uppercase\expandafter{\romannumeral3}, MTL\_\uppercase\expandafter{\romannumeral4}, NC-MTL), respectively. }\label{weight}
\end{figure}

\section{Experimental Results}\label{expem}

\subsection{Data acquisition and preprocessing}
In this study, data were taken from the Philadelphia Neurodevelopmental Cohort (PNC) \cite{pncdata1}, which is a collaborative study of child development between the Brain Behavior Laboratory at the University of Pennsylvania and the Center for Applied Genomics at the Children's Hospital of Philadelphia. The PNC contained nearly $900$ participants ($8\!-\!22$ years old) with multi-modal neuroimaging and genetics datasets. Our analyses were limited to $715$ subjects who underwent rs-fMRI scans and had minimal head movement with a mean frame-wise displacement being less than $0.25$ mm. The demographic characteristics of the subjects are shown in Table \ref{table2}. During the resting-state scan, subjects were instructed to stay awake, keep eyes open, fixate on the displayed crosshair, and remain still.

\begin{table}[!h]
\centering
\renewcommand\arraystretch{1.25}
\caption{Demographic characteristics of the subjects in this study; std denotes the standard deviation.}
\label{table2}
\setlength{\tabcolsep}{3mm}{
\begin{tabular}{ccc}
\hline
                                        & Male                        & Female                      \\ \hline
Number of subjects                      & $319$                         & $396$                         \\
Age (range; $\text{mean}\pm\text{std}$) & $8.58\!-\!21.75$                 & $8.67\!-\!22.58$                 \\
                                        & $15.23\pm3.14$ & $15.67\pm3.17$ \\ \hline
\end{tabular}}
\end{table}

\begin{figure}[!t]
 \centering
\includegraphics[width=0.9\columnwidth]{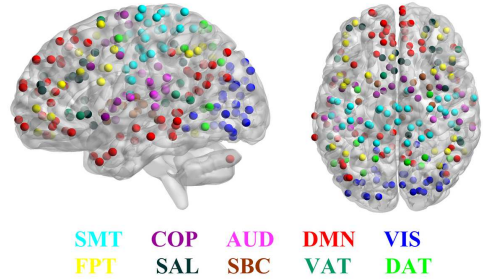}\\
\caption{The Power atlas with an a priori assignment of ROIs to different functional modules. ROIs of the same color belong to the same module and ROIs' colors indicate module memberships, where ROIs assigned to $10$ key functional modules were visualized and the others (assigned to cerebellar and unsorted) not.}\label{brain}
\end{figure}

All rs-fMRI datasets were acquired on the same $3$T Siemens TIM Trio whole-body scanner using a single-shot, interleaved multi-slice, gradient-echo, EPI sequence ($\text{TR/TE}=3000/32$ ms, $\text{flip angle}=90^{\circ}$, $\text{FOV}=192\times192$ $\text{mm}^2$, $\text{matrix}=64\times64$, $\text{resolution}\!=\!3\times3\times3$ $\text{mm}^3$, $124$ volumes). The scanning duration for each subject was about $6$ min, resulting in $124$ time points. Standard preprocessing procedures were applied to functional images using SPM12 (\url{www.fil.ion.ucl.ac.uk/spm/}), which include motion correction, co-registration, spatial normalization to standard MNI space, and temporal smoothing with a $3$ mm FWHM Gaussian kernel. The influences of head motion were regressed out, and functional time courses were further band-pass filtered with a passband of $0.01\!-\!0.1$ Hz. On the basis of the Power atlas \cite{power}, we segmented each subject's whole-brain into $264$ ROIs (modelled as $10$ mm diameter spheres), which spanned the cerebral cortex, subcortical structures, and the cerebellum. The majority of these ROIs ($227$ out of $264$) were assigned to $10$ pre-defined functional modules, i.e., sensory-motor network (SMT), default mode network (DMN), visual network (VIS), cingulo-opercular network (COP), fronto-parietal network (FPT), dorsal attention network (DAT), ventral attention network (VAT), auditory network (AUD), salience network (SAL), and subcortical network (SBC), which were utilized for localization analyses and visualized with BrainNet Viewer \cite{brainnetviewer} in Fig. \ref{brain}. A functional connectivity matrix ($264\times264$) was obtained for each subject by computing functional connectivity between any pair of ROIs. With removing duplicate functional connectivity, only the lower triangular portion of the symmetric functional connectivity matrix was unfurled into a feature vector of $34716$ functional connectivity for each subject in subsequent analysis.

\subsection{Comparison between univariate and multivariate functional connectivity for age prediction}
In this subsection, we utilized whole-brain functional connectivity (i.e., a total of $34716$ functional connectivity for each subject) to predict subjects' ages based on a linear support vector regression (SVR). For comparison, two different methods introduced in Section \ref{subsec2.1} were adopted to construct functional connectivity, i.e., dCor and pCor based functional connectivity, respectively. The SVRs (implemented in LIBSVM with default parameters \cite{ccchang}) were trained and tested using $5$-fold CV, and the 5-fold CV procedure was repeated $10$ times to reduce the effects of CV sampling bias and provide reliable performance. We reported the average prediction performance (mean $\!\pm\!$ std), which was quantified by both correlation coefficient (cc) and rmse between the predicted and observed ages of the subjects in the test sets over all runs of CV.

\begin{figure}[!t]
 \centering
\includegraphics[width=0.85\columnwidth]{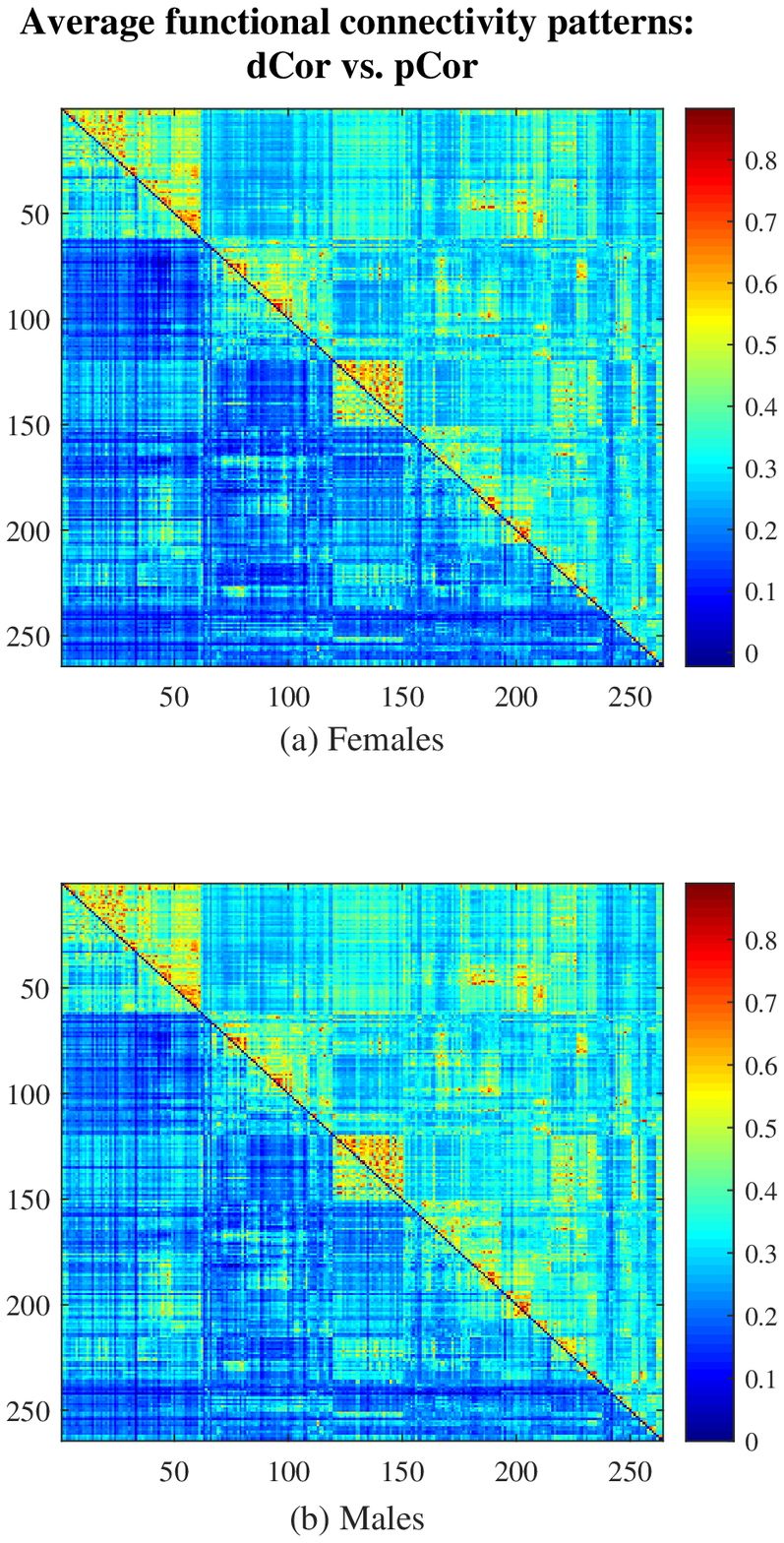}\\
\vspace{-0.65cm}
\caption{The average functional connectivity patterns estimated by dCor (upper triangle of a matrix heatmap) and pCor (lower triangle) across subjects for each gender group.}\label{FCN}
\end{figure}

\begin{figure}[!t]
    \centering
    \subfigure[Age prediction for females]{
        \includegraphics[width=0.85\columnwidth]{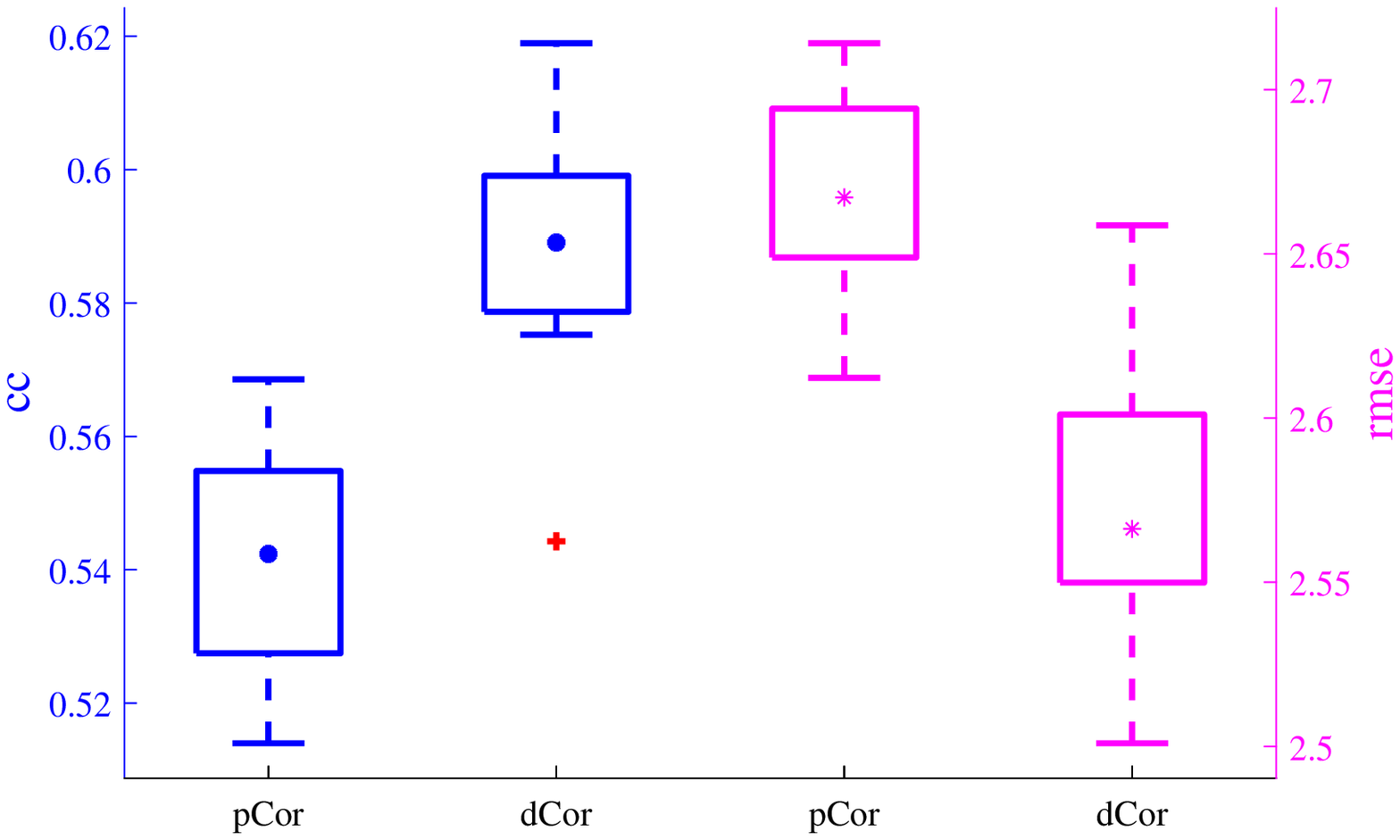}}
    \subfigure[Age prediction for males]{
    \includegraphics[width=0.85\columnwidth]{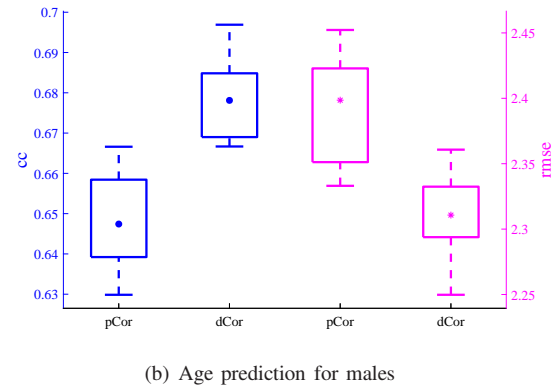}}
    \vspace{-0.1cm}
    \caption{The prediction performance in terms of cc and rmse for each gender group. Blue box plots exhibit cc results for the left $y$-axis, and magenta box plots exhibit rmse results for the right $y$-axis, where $\bullet$ and $\ast$ indicate the corresponding mean values.}
    \label{FCN2}
\end{figure}

Fig. \ref{FCN} illustrates the average dCor and pCor based functional connectivity patterns across subjects for each gender group. In Fig. \ref{FCN}, the average dCor based functional connectivity shown in the upper triangle of a matrix heatmap is clearly stronger than the average pCor based functional connectivity shown in the lower triangle. The age prediction performance for each gender group is presented in Fig. \ref{FCN2}. Specifically, for the female group, cc and rmse results using dCor based functional connectivity were $0.5891\pm0.0207$ and $2.5662\pm0.0459$, respectively, which were better than the corresponding ones (i.e., $0.5424\pm0.0169$ and $2.6672\pm0.0306$) using pCor based functional connectivity. Similarly, for the male group, the prediction results using dCor based functional connectivity were also better than those using pCor based functional connectivity, i.e., $0.6781\pm0.0103$ and $2.3107\pm0.0340$ vs. $0.6474\pm0.0118$ and $2.3986\pm0.0407$. This suggests that dCor based functional connectivity is more discriminative for age prediction than pCor based functional connectivity. By exploring spatial relations of voxel-wise time courses within each ROI, multivariate functional connectivity estimates (e.g., distance correlation) can provide more powerful information about individuals' unique brain organizations than univariate estimates. Therefore, in what follows we only focus on dCor based functional connectivity to jointly analyze age prediction tasks for both genders.

\begin{table*}[!t]
\centering
\renewcommand\arraystretch{1.75}
\caption{The comparison of regression performance of the male group and the female group by different predictive models.}
\label{t_result}
\setlength{\tabcolsep}{4mm}{
\begin{tabular}{ccccc}
\hline
\multirow{2}{*}{Model} & \multicolumn{2}{c}{Males}                            & \multicolumn{2}{c}{Females}     \\ \cline{2-5}
                       & cc ($\text{mean}\pm\text{std}$) & \multicolumn{1}{c|}{rmse ($\text{mean}\pm\text{std}$)} & cc ($\text{mean}\pm\text{std}$) & rmse ($\text{mean}\pm\text{std}$) \\ \hline
SVR                    & $0.6297\pm0.0191$ & \multicolumn{1}{c|}{$2.4615\pm0.0455$}   & $0.5119\pm0.0215$ & $2.7599\pm0.0449$   \\ \hline
MTL\_\uppercase\expandafter{\romannumeral1}                   & $0.6432\pm0.0102$ & \multicolumn{1}{c|}{$2.4239\pm0.0397$}   & $0.5140\pm0.0197$ & $2.7560\pm0.0433$   \\ \hline
MTL\_\uppercase\expandafter{\romannumeral2}                   & $0.6441\pm0.0195$ & \multicolumn{1}{c|}{$2.4080\pm0.0554$}   & $0.5210\pm0.0198$ & $2.7380\pm0.0424$   \\ \hline
MTL\_\uppercase\expandafter{\romannumeral3}                  & $0.6486\pm0.0083$ & \multicolumn{1}{c|}{$2.3958\pm0.0222$}   & $0.5364\pm0.0181$ & $2.6970\pm0.0382$   \\ \hline
MTL\_\uppercase\expandafter{\romannumeral4}                   & $0.6491\pm0.0183$ & \multicolumn{1}{c|}{$2.3918\pm0.0517$}   & $0.5362\pm0.0183$ & $2.6976\pm0.0386$   \\ \hline
NC-MTL                 & $0.6600\pm0.0096$ & \multicolumn{1}{c|}{$2.3632\pm0.0318$}   & $0.5452\pm0.0164$ & $2.6761\pm0.0358$   \\ \hline
\end{tabular}}
\end{table*}

\subsection{Results of the proposed NC-MTL for age prediction}
In this subsection, with the use of dCor based functional connectivity, we compared the age prediction performance of our NC-MTL model with five other predictive models, i.e., SVR for each gender group separately, and four MTL models (MTL\_\uppercase\expandafter{\romannumeral1}, MTL\_\uppercase\expandafter{\romannumeral2}, MTL\_\uppercase\expandafter{\romannumeral3}, MTL\_\uppercase\expandafter{\romannumeral4}) as mentioned before. We used $10$ times $5$-fold nested CV to tune the hyperparameters as well as to obtain the best average performance in all experiments. All regularization parameters (also called hyperparameters) in the five MTL models were chosen by a grid search within their respective ranges; that is, $\alpha, \beta\in\{10^{-4},10^{-3},10^{-2},10^{-1},1,10\}$. Prior to training the predictive models, simple feature filtering was conducted. More specifically, we discarded the dCor based functional connectivity features for which the $p$-values of the correlation with ages of males and females in the training set were both greater than or equal to $0.01$. For each gender group, the remaining features of training subjects were normalized to have zero mean and unit norm, and the mean and norm values of training subjects were used to normalize the corresponding features of testing subjects. We performed the mean-centering on ages of training subjects and then used the mean age value of training subjects to normalize ages of testing subjects.

\begin{figure}[!t]
 \centering
\includegraphics[width=0.8\columnwidth]{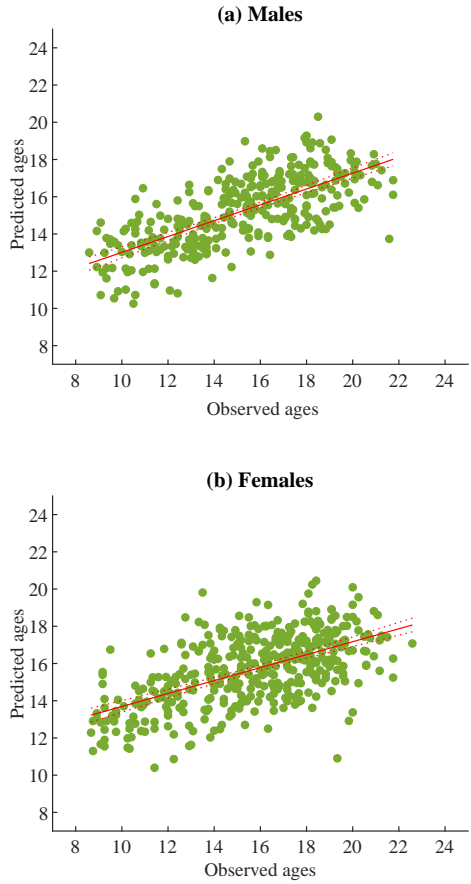}\\
 \vspace{-0.75cm}
\caption{The two scatter plots illustrate the relationships between the predicted and observed ages of males and females, respectively, where the predicted ages were obtained by the proposed NC-MTL model. Each green dot represents one subject. Each red solid line represents the best-fit line of the green dots, and its $95\%$ confidence interval is indicated by two dashed lines.}\label{relation_predi_obser}
\end{figure}

The detailed age prediction results are summarized in Table \ref{t_result}. The accuracy of the proposed NC-MTL model was always superior to those of other predictive models, indicating that our NC-MTL model had better prediction performance. It suggests that the composite regularizer by combining $\ell_{2,1-2}$ and $\ell_{1-2}$ regularization terms, introduced in our NC-MTL model, was more effective in identifying discriminative features associated with ages through selecting both common and gender-specific features. Moreover, as shown in Table \ref{t_result}, the five MTL models all achieved better prediction performance than the STL model (i.e., SVR), which demonstrates that joint analysis of multiple tasks, while exploiting commonalities and/or differences across tasks, can result in improved prediction accuracy, compared to learning these tasks independently. For the proposed NC-MTL model, we present the relationships between the predicted and observed ages of males and females in Fig. \ref{relation_predi_obser}, respectively.

\begin{figure}[!h]
 \centering
 \hspace*{-0.85cm}
\includegraphics[width=1.15\columnwidth]{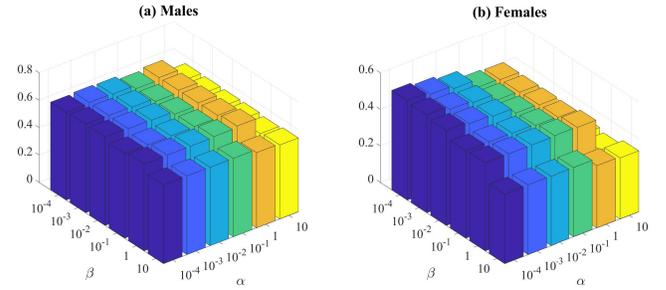}\\
 \vspace{-0.6cm}
\caption{The cc results of both genders based on the proposed NC-MTL model with different values of $\alpha$ and $\beta$. }\label{parameter_effect}
\end{figure}

In the objective function (\ref{qingtian}) of our NC-MTL model, there are two regularization parameters (i.e., $\alpha$ and $\beta$). They balance the relative contributions of the common and task-specific feature selection, respectively. We then studied the effect of these regularization parameters on the age prediction performance. As shown in Fig. \ref{parameter_effect}, the parameters $\alpha$ and $\beta$ were combined to obtain the age prediction performance of the proposed NC-MTL model, which fluctuates when changing the values of the parameters.

\subsection{Discriminative functional connectivity and gender differences detected by the proposed NC-MTL}
In this subsection, based on the proposed NC-MTL model, we investigated the most discriminative functional connections (functional connectivity features) with potential biological significance relevant to gender differences in brain development. Specifically, the proposed NC-MTL model in (\ref{qingtian}) generated two weight vectors (i.e., $\textbf{w}_1$ and $\textbf{w}_2$, one for each gender group) of functional connectivity features. With respect to each gender group, we averaged the absolute values of the weights of each feature over all runs of CV as the weight of the corresponding functional connectivity. The larger the weight of the functional connectivity feature is, the more discriminative the functional connectivity feature is.

\begin{figure}[!t]
    \centering
    \hspace*{-0.75cm}
    \subfigure[]{
        \includegraphics[width=0.45\columnwidth]{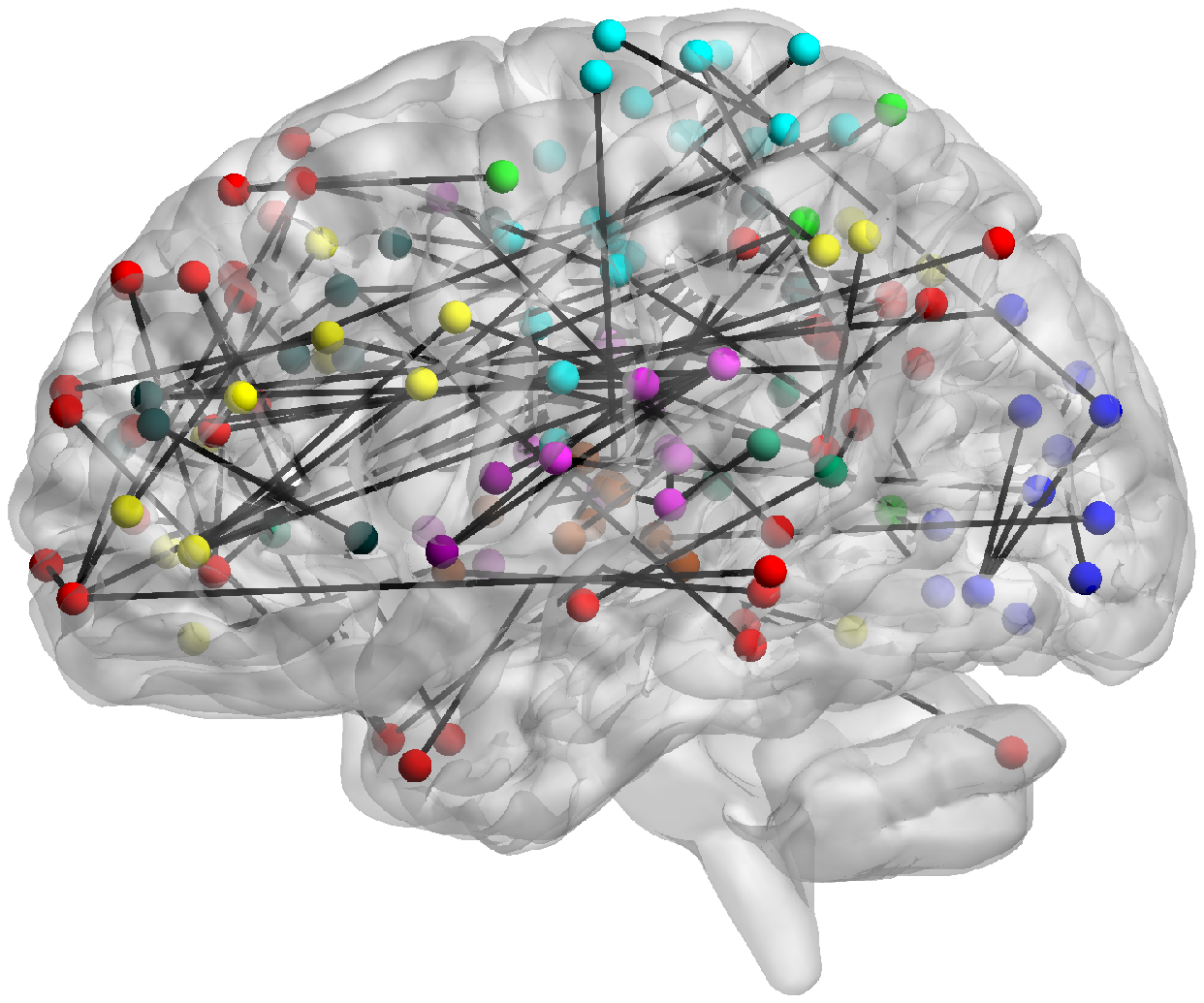}}
        \hspace*{-0.72cm}
    \subfigure[]{
        \includegraphics[width=0.65\columnwidth]{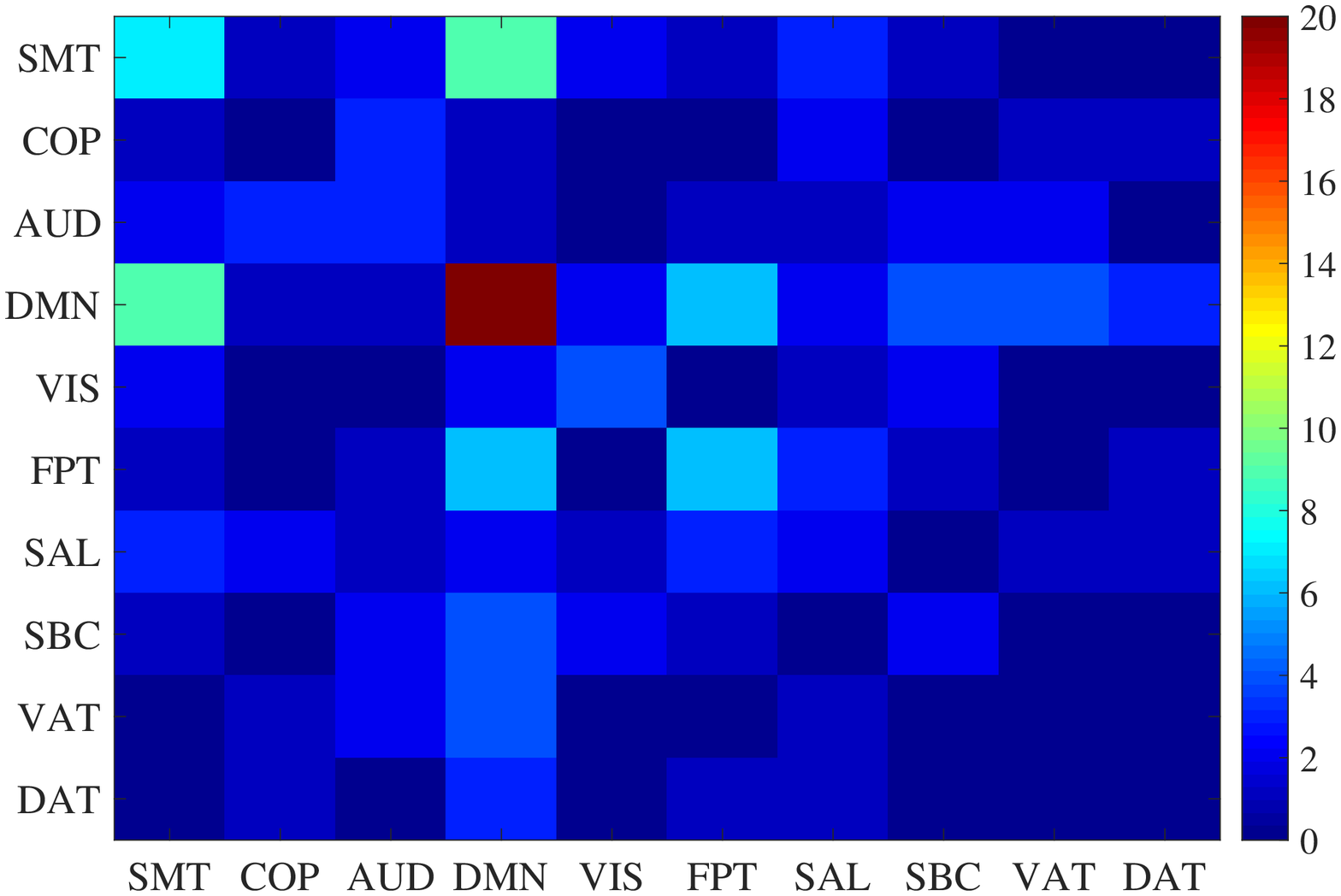}}
        \hspace*{-0.75cm}
    \subfigure[]{
        \includegraphics[width=0.45\columnwidth]{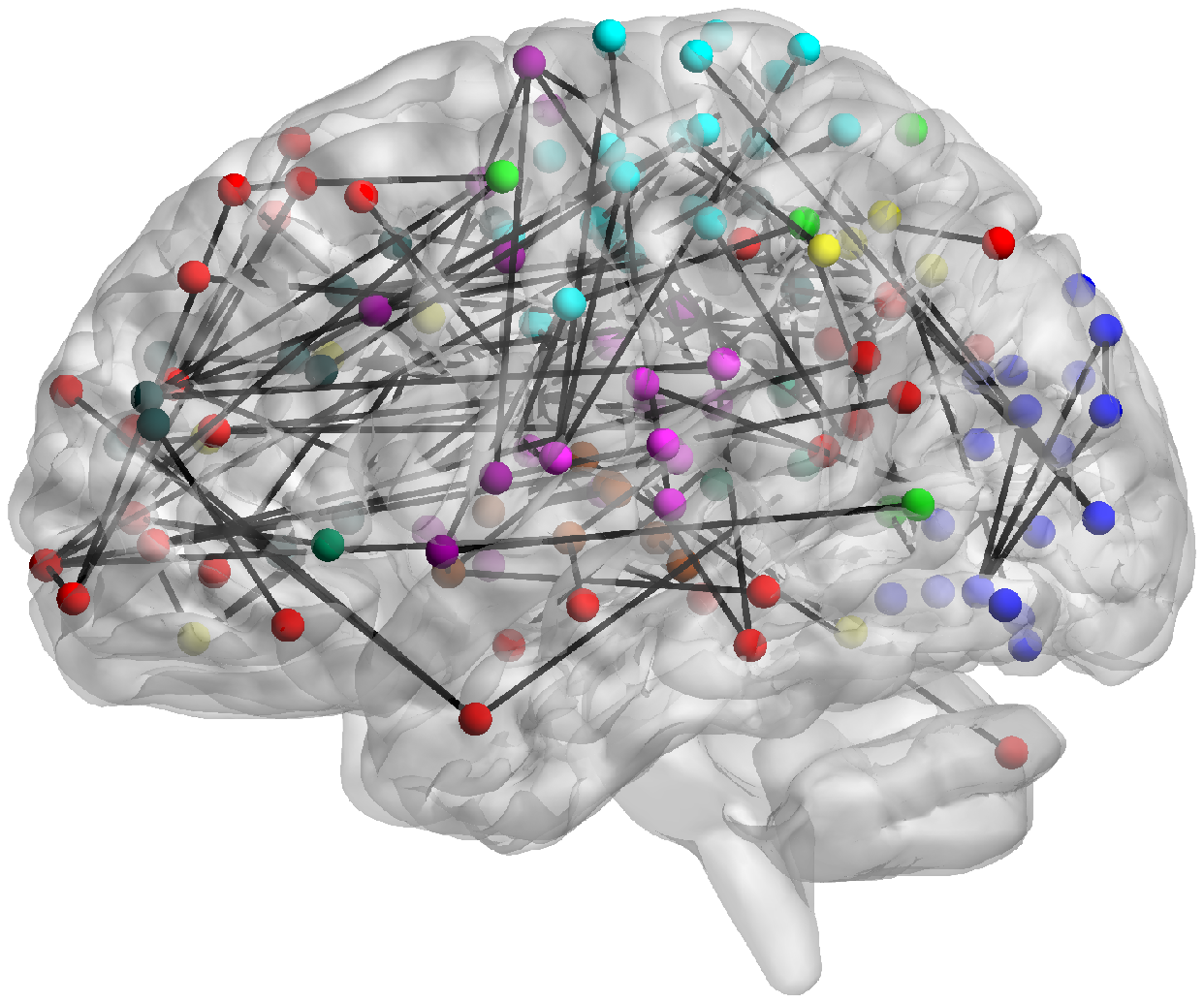}}
        \hspace*{-0.72cm}
    \subfigure[]{
    \includegraphics[width=0.65\columnwidth]{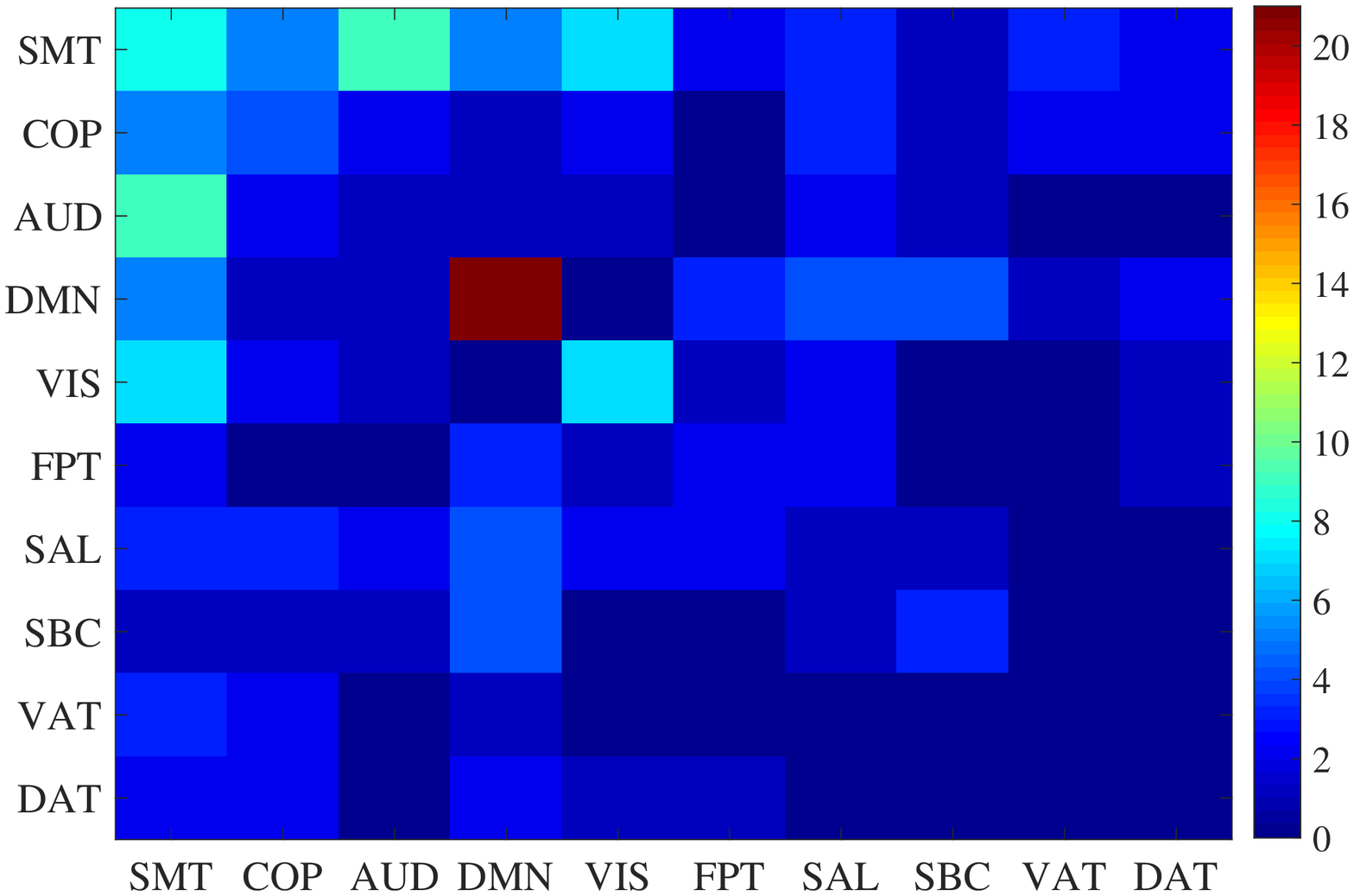}}
    \caption{The visualization of the most discriminative (among 150) age-related functional connections between and within the $10$ functional modules for each gender group, i.e., (a)-(b) males and (c)-(d) females. The left are brain plots showing sagittal views of the functional graph in anatomical space, where node colors indicate module membership. The right are matrix plots showing the total numbers of within- and between-module connections.}
    \label{brain_edge}
\end{figure}

For ease of visualization, we identified the top $150$ most discriminant age-related functional connections for each gender group, and Fig. \ref{brain_edge} only shows the most discriminant within- and between-module functional connections for the $10$ pre-defined functional modules. As shown in Fig. \ref{brain_edge}, SMT, DMN, VIS, and FPT are important functional modules detected for both genders. The numbers of identified functional connections between SMT and DMN, between FPT and DMN, and within FPT are larger for males. The numbers of identified functional connections between SMT and AUD, within VIS, and between SMT and VIS are larger for females. Functional brain activity spanning the frontoparietal regions were involved in comparing heading direction \cite{finding1}, and functional connections between the right FPT and DMN were increased in better navigators \cite{finding2}. For females higher connectivity existed between sensory and attention systems, while for males higher connectivity between sensory, motor, and default mode systems were observed \cite{finding3}. Recent evidence indicates that functional connectivity patterns of the auditory system and many other (e.g., visual and motor) brain systems were related to language-related activation \cite{finding4}.
Therefore, these findings in this paper were consistent with the previous results that males have better spatial orientation and motor coordination skills, and females have better visual language and verbal working memory skills.

\subsection{Limitations and future work}
In this paper, we estimated functional connectivity between ROIs using distance correlation rather than Pearson's correlation. Distance correlation is a multivariate statistical method, which is able to measure both linear and nonlinear dependence between ROIs, and hence captures more complex information. However, like Pearson's correlation, distance correlation cannot exclude the effects of several other controlling or confounding ROIs when computing pairwise correlations. Therefore, in our follow-up study, it is interesting to measure functional connectivity by partial distance correlation \cite{szekely2014,jianfang2018}, which is an extension of distance correlation, and can calculate conditional dependence between ROIs. Furthermore, the proposed NC-MTL model achieved satisfactory prediction performance, but we can further improve it in our future work. For example, in our NC-MTL model, we can impose additional constraints that effectively utilize different pieces of information inherent in the data, including feature-feature relation, label-label relation, and subject-subject relation \cite{zhuzhu2017}. As deep neural networks have recently received growing attention and shown outstanding performance in various applications, it is also interesting to extend the composite regularizer in our NC-MTL model into a multi-task deep learning framework. On the other hand, it will be important to apply our NC-MTL model to evaluate differences in brain functional connectivity patterns across different populations, e.g., disease conditions, or developmental stages in behavior and cognition.

\section{Conclusion}\label{conclusion}
In this paper, we first demonstrated that multivariate functional connectivity estimates can provide more powerful information between ROIs than univariate functional connectivity estimates. The experimental results on the PNC data showed that dCor based functional connectivity better predicted individuals' ages than pCor based functional connectivity. Next, we proposed a novel NC-MTL model by introducing a composite regularizer that combines the $\ell_{2,1-2}$ and $\ell_{1-2}$ terms, which are improved alternatives to the classical $\ell_{2,1}$ and $\ell_{1}$, respectively; as a result, it promises improved extraction of common and task-specific features. Results showed improved performance of the proposed NC-MTL model over several competing ones for predicting ages from functional connectivity patterns using rs-fMRI of the PNC, where age prediction for each gender group was treated as one task. In addition, we detected both common and gender-specific age-related functional connectivity patterns to characterize the effects of gender and age on brain development.

\end{document}